\documentstyle[preprint,aps,psfig,12pt]{revtex}

\begin{document}

\tightenlines

\title{Lindstedt Series Solutions of the Fermi-Pasta-Ulam Lattice }

\author{David C. Dooling and James E. Hammerberg}

\address{
Los Alamos National Laboratory,
\\
Los Alamos, New Mexico 87545
\\}

\maketitle

\begin{abstract}
We apply the Lindstedt method to the one dimensional Fermi-Pasta-Ulam $\beta$ lattice to find
fully general solutions to the complete set of equations of motion.
 The pertubative scheme employed uses
$\epsilon$ as the expansion parameter, where $\epsilon$ is the coefficient of the quartic coupling between nearest neighbors.
We compare our non-secular perturbative solutions
to numerical solutions and find striking agreement. 
\end{abstract}

\section{Normal Mode Representation}
No known solution for a coupled set of Duffing oscillators exists, with the exception
of specific cases.
The Duffing lattice is perhaps more well known as the FPU $\beta$-system, one of the
systems studied by Fermi, Pasta, and Ulam in the $1950s$
\cite{Bivins} to numerically investigate the approach to thermal equilibrium in nonlinear systems.
FPU found that the expected equipartition of energy among the linear normal modes did not occur, and
that periodic, non-ergodic motion persisted even in the presence of the nonlinear mixing terms.
Subsequent investigations, both numerical and analytic, have shown that different parameter regions
specified by the total energy and the strength of the nonlinear terms as specified by the parameter
$\epsilon$ lead to markedly different types of behaviour \cite{Pettini}, with the expected ergodicity emerging with
increasing energy and $\epsilon$.
But the observation of the periodic motion in the orginal FPU $\beta$-system computer experiments
lends credence to the possibility of finding acceptable pertubative solutions in certain
parameter regions.
That no exact solution exsists for a large $N$ - body lattice modulo some
exceptional cases \cite{Kosevich} motivates us to develop pertubative solutions to systems of Duffing
oscillators.

Several pertubation schemes as applied to the Duffing oscillator have been investigated
\cite{Mandal}, but the methods are applicable only in the weakly nonlinear regime and also
require the time $t$ to be small in order to be rigorously trustworthy.
We therefore apply a different perturbative procedure called the Lindstedt series.
We note that similar perturbation schemes
that  avoid secular
terms
have been proposed \cite{Sholl,Ford1,Ford2,Jackson1,Jackson2,Sholl2,Christie}
 in the past.
In particular, we note that the method presented here is a more general variation
of the so-called shifted frequency perturbation scheme, as described in \cite{Christie}.
The Lindstedt method is a systematic procedure to compute formal power series expansions of quasi-periodic solutions \cite{Rand}.
We consider an anharmonic bath of $N$ degrees of freedom in one dimension with fixed endoints 
$(q_{0} = q_{N+1} = 0)$, (we consider these boundary conditions
as opposed to periodic boundary conditions so as to avoid zero modes, which are 
problematic when using the lattice as a bath coupled to a test particle in the
standard derivation of the generalized Langevin equation) as described by the Hamiltonian:

\begin{equation}
H = \frac{1}{2} \sum_{i=1}^{N} p_{i}^{2} + \frac{1}{2} \sum_{i=1}^{N+1} \left( q_{i} - q_{i-1} \right)^{2} 
+ \frac{\epsilon}{4} \sum_{i=1}^{N+1} \left( q_{i} - q_{i-1} \right)^{4} \,.
\end{equation}

The normal coordinates ${Q_{i}}$ are related to the site coordinates ${q_{j}}$ through the transformations
\begin{equation}
q_{i} = \sum_{j=1}^{N} A_{ij} Q_{j}
\end{equation}
and
\begin{equation}
p_{i} = \sum_{j=1}^{N} A_{ij} P_{j}
\end{equation}
where
\begin{equation}
A_{ij} = \sqrt{ \frac{2}{N+1} } \sin \left( \frac{ \pi i j}{N+1} \right) \,.
\end{equation}
By means of this transformation, the harmonic part of the Hamiltonian is decomposed into a sum of
independent normal modes:
\begin{equation}
H_{0} = \sum_{k=1}^{N} \frac{1}{2}
\left( P_{k}^{2} + \omega_{k}^{2} Q_{k}^{2} \right),
\end{equation}
where
\begin{equation}
\omega_{k}^{2} = 4 \sin^{2} \left( \frac{ \pi k}{2(N+1)} \right) \,.
\end{equation}

 We have the following compact expresssion for $H$ expressed in terms of the
linear normal modes \cite{Shinohara,Sholl,Shin2}:

\begin{eqnarray*}
H  & = & \frac{1}{2} \sum_{k=1}^{N} \left( P_{k}^{2} + \omega_{k}^{2} Q_{k}^{2} \right) \\
 & + & \frac{\epsilon}{8 \left( N + 1 \right)} \sum_{k=1}^{N} \sum_{l=1}^{N} \sum_{m=1}^{N} \sum_{n=1}^{N}
\omega_{k} \omega_{l} \omega_{m} \omega_{n} C_{klmn} Q_{k} Q_{l} Q_{m} Q_{n},
\end{eqnarray*}
where the coupling coefficients $C_{klmn}$ are given by
\begin{eqnarray*}
C_{klmn} & = & \Delta_{k+l+m+n} + \Delta_{k-l+m+n} + \Delta_{k+l-m+n} \\
 & + & \Delta_{k+l+m-n} + \Delta_{k-l-m+n} + \Delta_{k+l-m-n} \\
 & + & \Delta_{k-l+m-n} + \Delta_{k-l-m-n},
\end{eqnarray*}
with $\Delta_{r}$ defined by
\begin{equation}
\Delta_{r} = \left\{ \begin{array}{cl}
1 & \mbox{for $r=0$}, \\
-1 &  \mbox{for $r = \pm 2 \left(N + 1 \right)$}, \\
0 &  \mbox{otherwise} \end{array} \right. .
\end{equation}

\section{Lindstedt Method}
We seek solutions to the equations of motion for $Q_{k}(t)$,
\begin{equation}
\frac{ \mbox{d}^{2} Q_{k}}{\mbox{d} t^{2}}  + \omega_{k}^{2} Q_{k}  =  
-\frac{\epsilon }{2(N+1)} \sum_{l,m,n} \omega_{k} \omega_{l} \omega_{m} \omega_{n} C_{klmn} Q_{l} Q_{m} Q_{n} 
\label{YO}
\end{equation}
where $\epsilon$ is a small parameter $(\epsilon < 1)$.
A hallmark of nonlinear, periodic behaviour is the amplitude dependence of the frequency.
To capture this relationship, for each normal mode $Q_{k}$
 we define a new stretched temporal variable
$\tau_{k} = \beta_{k} t$
where to first order in $\epsilon$ we define
$\beta_{k} = 1 + \epsilon \rho_{k}$ and the quantity $\rho_{k}$ is to be determined.
The Lindstedt method can in principle be expanded to all powers in $\epsilon$ \cite{Rand},
although the convergence properties for arbitrary $N$ and $\epsilon$ are not known.
In this work, we will only retain expressions up to first order in $\epsilon$ in
the quantities $\{ \beta_{k} \}$.

The Lindstedt method then makes the assumption that we can express $Q_{k}(\tau_{k})$ in the form
\begin{equation}
Q_{k}   \approx    Q_{k,0} + \epsilon Q_{k,1}.
\label{BigQ}
\end{equation} 

The zeroth order equations are $(\epsilon = 0)$:
\begin{equation}
\frac{\mbox{d}^{2} Q_{k,0}}{\mbox{d} \tau_{k}^{2}} + \omega_{k}^{2} Q_{k,0} = 0,
\end{equation}
with solutions
\begin{equation}
Q_{k,0} (\tau_{k}) = Q_{k} (0) \cos (\omega_{k} \tau_{k})
 + \frac{ Q^{\prime}_{k}(0)}{\omega_{k}} \sin (\omega_{k} \tau_{k}). 
\label{zerosol}
\end{equation}

The order $\epsilon$ equations are given by
\begin{eqnarray}
\frac{ \mbox{d}^{2} Q_{k,1}}{\mbox{d} \tau_{k}^{2}} 
 + \omega_{k}^{2} Q_{k,1} & = & 2 \rho_{k} \omega_{k}^{2} Q_{k,0}(\tau_{k}) 
\nonumber \\
& &  -\frac{\omega_{k}}{2(N+1)} \sum_{l,m,n=1}^{N} \omega_{l} \omega_{m} \omega_{n} C_{klmn}
 Q_{l,0}(\tau_{l}) 
Q_{m,0}(\tau_{m})
Q_{n,0}(\tau_{n}) \,. 
\label{CUEONE}
\end{eqnarray}
or, in terms of the original time variable $t$:
\begin{eqnarray}
\frac{ \mbox{d}^{2} Q_{k,1}}{\mbox{d} t^{2}} 
 + \beta_{k}^{2} \omega_{k}^{2} Q_{k,1} & = &
 2 \rho_{k} \beta_{k}^{2}  \omega_{k}^{2} Q_{k,0}(\beta_{k} t) \nonumber \\ 
& &  -\frac{ \beta_{k}^{2} \omega_{k}}{2(N+1)}
 \sum_{l,m,n=1}^{N} \omega_{l} \omega_{m} \omega_{n} C_{klmn}
 Q_{l,0}(\beta_{l} \omega_{l} t ) 
Q_{m,0}(\beta_{m} \omega_{m} t)
Q_{n,0}(\beta_{n} \omega_{n} t) \,. 
\label{WOW}
\end{eqnarray}

In order to apply the Lindstedt method, we need to identify all of the potential resonant
 driving terms \cite{Rand}.
Terms with the
 arguments $(\beta_{l} \omega_{l} +\beta_{m}  \omega_{m} +\beta_{n} \omega_{n})t$
 will never provide resonance terms.
Using the symmetry properties of the coupling coefficients $C_{klmn}$, we 
arrive at the following resonant forcing terms in the equations of motion for the $Q_{k,1}$:
\begin{eqnarray}
\frac{\mbox{d}^{2} Q_{k,1}}{\mbox{d} t^{2}} +\beta_{k}^{2}  \omega_{k}^{2} Q_{k,1}  &=&
  2 \rho_{k} \beta_{k}^{2}  \omega_{k}^{2}
\left[Q_{k}(0)\cos (\beta_{k}\omega_{k} t)+\frac{\dot{Q}_{k}(0)}{\beta_{k}\omega_{k}}
\sin (\beta_{k} \omega_{k} t)\right] 
\nonumber\\
&&
-  \frac{\beta_{k}^{2} \omega^{4}_{k}}{8(N+1)}  C_{kkkk} \times \Bigg\{
\left[3Q_{k}^{3}(0)+3\left(\frac{\dot{Q}_{k}(0)}{\beta_{k} \omega_{k}}\right)^{2} Q_{k}(0) \right]
\cos(\omega_{k} \beta_{k}  t) 
\nonumber\\
&& \qquad\qquad
+ \Bigg[ 3\left(\frac{\dot{Q}_{k}(0)}{\beta_{k} \omega_{k}}\right)^{3}
+3Q_{k}^{2}(0) \frac{\dot{Q}_{k}(0)}{\beta_{k} \omega_{k}} \Bigg]
 \sin(\omega_{k} \beta_{k}  t) \Bigg\}
\nonumber\\
 &&  
-\frac{3 \beta_{k}^{2} \omega_{k}^{2}}{8(N+1)} \sum_{m \neq k}^{N} \omega_{m}^{2} C_{kkmm}   \Bigg\{ 
\Bigg[2Q_{k}(0)Q_{m}^{2}(0)+2Q_{k}(0)\left(\frac{\dot{Q}_{m}(0)}{\beta_{m} \omega_{m}}\right)^{2} \Bigg]
\cos(\omega_{k} \beta_{k}  t)
\nonumber\\
&& \qquad\qquad
+
\Bigg[2Q_{m}^{2}(0)\frac{\dot{Q}_{k}(0)}{\beta_{k} \omega_{k}}
+2\frac{\dot{Q}_{k}(0)}{\beta_{k} \omega_{k}} 
 \left(\frac{\dot{Q}_{m}(0)}{\beta_{m} \omega_{m}}\right)^{2} \bigg]
 \sin(\beta_{k} \omega_{k} t) \Bigg\} \,. 
\nonumber\\
&& \qquad\qquad 
+ \mbox{non-resonant} \, \, \mbox{terms}. 
\label{WOW2}
\end{eqnarray}

Setting to zero the coefficients mutiplying
 $\sin (\beta_{k} \omega_{k} t)$ and $\cos(\beta_{k}  \omega_{k} t)$ yields the following
system of coupled nonlinear algebraic equations for 
 the parameters $\{ \rho_{k} \}$:
\begin{eqnarray}
\rho_{k} & = &
\frac{3 \omega_{k}^{2}}{16(N+1)} C_{kkkk}
\Bigg\{Q_{k}^{2}(0) + 
\left(\frac{\dot{Q}_{k}(0)}{(1+\epsilon \rho_{k})\omega_{k}}\right)^{2} \Bigg\}
\nonumber \\
& + & \frac{3}{8(N+1)} \sum_{m \neq k}^{N} \omega_{m}^{2} C_{kkmm}
\Bigg\{Q_{m}^{2}(0) + 
\left(\frac{\dot{Q}_{m}(0)}{(1 + \epsilon \rho_{m})\omega_{m}}\right)^{2} \Bigg\}.
\label{rhokay}
\end{eqnarray}
In principle, one may solve this system of equations for the
set $\{ \rho_{k} \}$.
Having obtained $\{ \rho_{k} \}$, one then substitutes the expressions
given by Eqn.~(\ref{zerosol}) into the right hand side of Eqn.~(\ref{WOW})
and uses the harmonic oscillator Green's function to solve for 
the set $\{ Q_{k,1} \}$.
In the present work, for the sake of simplicity. we restrict our 
attention to corrections to the harmonic frequencies $\{ \omega_{k} \}$
that are first order in $\epsilon$; i.e., we consider
$\beta_{k} = 1 + \epsilon \rho_{k}$ where $\rho_{k}$ is independent of
$\epsilon$.
If one wishes to compute first order corrections to the
$\{ \omega_{k} \}$, then $\epsilon$ may be set to zero
in the right hand side of Eqn.~(\ref{rhokay}), and one obtains:

\begin{eqnarray}
\rho_{k} & = &  
\frac{3\omega_{k}^{2}}{16(N+1)}C_{kkkk}
\left\{Q_{k}^{2}(0)+\left(\frac{\dot{Q_{k}}(0)}{\omega_{k}}\right)^{2} \right\}
\nonumber  \\
& + & \frac{3 }{8(N+1)} \sum_{m \neq k}^{N} \omega_{m}^{2} C_{kkmm}
\left\{Q_{m}^{2}(0) + \left(\frac{\dot{Q_{m}}(0)}{\omega_{m}}\right)^{2} \right\},
\label{hewrho}
\end{eqnarray}
and $\beta_{k} = 1 + \epsilon \rho_{k}$.

Therefore the zeroth order normal mode $Q_{k,0}(t)$ is modified  as follows:
\begin{eqnarray}
Q_{k,0}(t) &  = &  Q_{k}(0) \cos(\omega_{k} \beta_{k} t)
 + \frac{\dot{Q}_{k}(0)}{\omega_{k}\beta_{k}} \sin (\omega_{k} \beta_{k} t) \label{erst} \\
\end{eqnarray}
where
\begin{eqnarray}
\beta_{k} & = & 1 +
\frac{3\epsilon\omega_{k}^{2}}{16(N+1)}C_{kkkk}
\left\{Q_{k}^{2}(0)+\left(\frac{\dot{Q_{k}}(0)}{\omega_{k}}\right)^{2} \right\} \nonumber \\
& + & \frac{3 \epsilon}{8(N+1)} \sum_{m \neq k}^{N} \omega_{m}^{2} C_{kkmm}
\left\{Q_{m}^{2}(0) + \left(\frac{\dot{Q_{m}}(0)}{\omega_{m}}\right)^{2} \right\}.
\end{eqnarray}

With this expression for the parameters $\beta_{k}$, where $\rho_{k}$ is 
independent of $\epsilon$, we have not removed resonant driving
terms from Eqn.~(\ref{WOW}), but rather from the evolution equation for
$Q_{k,1}(t)$ obtained by setting all of the set $\{ \beta_{k} \}$
equal to unity in Eqn.~(\ref{WOW}).
With the $\{ \rho_{k} \}$ given by Eqn.~(\ref{hewrho}), we have removed
resonant driving terms from the equations:
\begin{eqnarray}
\frac{ \mbox{d}^{2} Q_{k,1}}{\mbox{d} t^{2}} 
 + \omega_{k}^{2} Q_{k,1} & = & 2 \rho_{k} \omega_{k}^{2} Q_{k,0}(\omega_{k} t) 
\nonumber \\
& &  -\frac{\omega_{k}}{2(N+1)} \sum_{l,m,n=1}^{N} \omega_{l} \omega_{m} \omega_{n} C_{klmn}
 Q_{l,0}(\omega_{l} t) 
Q_{m,0}(\omega_{m} t)
Q_{n,0}(\omega_{n} t) \,. 
\label{WOW2}
\end{eqnarray}
Therefore, consistency requires that we solve the
Eqns.~(\ref{WOW2}) to determine the $Q_{k,1}(t)$.

 Having solved explicitly for the set ${ Q_{k,0}(t) }$ in terms 
of initial conditions, we may now solve for
the set ${ Q_{k,1}(t)}$.
The equation of motion for $Q_{k,1}(t)$ then becomes
\begin{equation}
\frac{\mbox{d}^{2} Q_{k,1}}{\mbox{d} t^{2}} + \omega_{k}^{2} Q_{k,1}  =    
 -\frac{\omega_{k}}{2(N+1)} \sum_{ \langle l,m,n=1 \rangle }^{N}
 \omega_{l} \omega_{m} \omega_{n} C_{klmn}
 Q_{l,0}(\omega_{l} t)Q_{m,0}(\omega_{m} t)Q_{n,0}(\omega_{n} t),
\end{equation}
where the brackets $\langle l,m,n=1 \rangle$ denote a restricted sum.
All of the amplitude dependence of the frequencies is carried by the $\{ Q_{k,0} \}$ 
when we only work to first order in $\epsilon$ in the $\{ \beta_{k} \}$.

The requirement that no secular terms appear on the right hand side, the same
 requirement that fixed the parameters $\rho_{k}$ and the set ${ \beta_{k} }$, 
has also guaranteed that there are no resonant driving terms in the above equation for $Q_{k,1}(t)$.
This requirement of no resonant terms has thus transformed the orginal sum into a restricted sum.
In the restricted sum, $\sum_{\langle l,m,n=1 \rangle}^{N}$, we discard all resonant driving terms.
A fixed set of indices ${l,m,n}$ will produce eight different harmonic source terms.
The meaning of the restricted sum is that we only keep the harmonic driving terms where the arguments do not
equal $\pm \omega_{k}$. 
The resonant terms have already been accounted for in our definition of $\rho_{k}$.

 The full solution $Q_{k}(t)$ is parametrized by
 two arbitrary constants determined from initial conditions, which
implies that the solution for $Q_{k,1}(t)$ will contain no arbitrary constants.
 Therefore we are only interested in the particular solution for $Q_{k,1}(t)$.

Rewriting the equation of motion as
\begin{eqnarray}
\frac{\mbox{d}^{2} Q_{k,1}}{\mbox{d} t^{2}} &+& \omega_{k}^{2} Q_{k,1}(t)  =
\nonumber\\
&&  
-\frac{\omega_{k}}{8(N+1)} \sum_{ \langle l,m,n=1 \rangle}^{N} \omega_{l} \omega_{m} \omega_{n} C_{klmn}
\Bigg\{
   \Bigg[ Q_{l}(0)Q_{m}(0)Q_{n}(0) +\frac{Q_{l}(0)\dot{Q}_{m}(0)\dot{Q}_{n}(0)}{\omega_{m}\omega_{n}} 
\nonumber\\
&& \qquad\qquad
 + \frac{Q_{m}(0)\dot{Q}_{l}(0)\dot{Q}_{n}(0)}{\omega_{l}\omega_{n}} 
 - \frac{Q_{n}(0)\dot{Q}_{m}(0)\dot{Q}_{l}(0)}{\omega_{m}\omega_{l}}  \Bigg]
\cos(( \omega_{l} +  \omega_{m} -  \omega_{n})t) 
\nonumber\\
&& \qquad\qquad
 + 
\Bigg[ Q_{l}(0)Q_{m}(0)Q_{n}(0) -\frac{Q_{l}(0)\dot{Q}_{m}(0)\dot{Q}_{n}(0)}{\omega_{m}\omega_{n}}
\nonumber\\
&& \qquad\qquad
 - \frac{Q_{m}(0)\dot{Q}_{l}(0)\dot{Q}_{n}(0)}{\omega_{l}\omega_{n}}
 - \frac{Q_{n}(0)\dot{Q}_{m}(0)\dot{Q}_{l}(0)}{\omega_{m}\omega_{l}}  \Bigg]
\cos(( \omega_{l} + \omega_{m} + \omega_{n})t) 
\nonumber\\
&&\qquad\qquad
\Bigg[ Q_{l}(0)Q_{m}(0)Q_{n}(0) -\frac{Q_{l}(0)\dot{Q}_{m}(0)\dot{Q}_{n}(0)}{\omega_{m}\omega_{n}} 
\nonumber\\
&&\qquad\qquad
 + \frac{Q_{m}(0)\dot{Q}_{l}(0)\dot{Q}_{n}(0)}{\omega_{l}\omega_{n}}
 + \frac{Q_{n}(0)\dot{Q}_{m}(0)\dot{Q}_{l}(0)}{\omega_{m}\omega_{l}} \Bigg]
\cos((\omega_{l} - \omega_{m} - \omega_{n})t)
\nonumber\\
&&\qquad\qquad
\Bigg[ Q_{l}(0)Q_{m}(0)Q_{n}(0) +\frac{Q_{l}(0)\dot{Q}_{m}(0)\dot{Q}_{n}(0)}{\omega_{m}\omega_{n}} 
\nonumber\\
&&\qquad\qquad
 - \frac{Q_{m}(0)\dot{Q}_{l}(0)\dot{Q}_{n}(0)}{\omega_{l}\omega_{n}}
 + \frac{Q_{n}(0)\dot{Q}_{m}(0)\dot{Q}_{l}(0)}{\omega_{m}\omega_{l}} \Bigg]
\cos((\omega_{l}-\omega_{m}+\omega_{n})t) 
\nonumber\\
&&\qquad\qquad
\Bigg[ \frac{\dot{Q}_{l}(0)\dot{Q}_{m}(0)\dot{Q}_{n}(0)}{\omega_{l}\omega_{m}\omega_{n}} 
+ \frac{Q_{m}(0)Q_{n}(0)\dot{Q}_{l}(0)}{\omega_{l}} 
\nonumber\\
&&\qquad\qquad
+ \frac{Q_{l}(0)Q_{n}(0)\dot{Q}_{m}(0)}{\omega_{m}} 
- \frac{Q_{l}(0)Q_{m}(0)\dot{Q}_{n}(0)}{\omega_{n}} \Bigg]
\sin((\omega_{l} + \omega_{m} - \omega_{n})t) 
\nonumber\\
&&\qquad\qquad
\Bigg[- \frac{\dot{Q}_{l}(0)\dot{Q}_{m}(0)\dot{Q}_{n}(0)}{\omega_{l}\omega_{m}\omega_{n}} 
+ \frac{Q_{m}(0)Q_{n}(0)\dot{Q}_{l}(0)}{\omega_{l}} 
\nonumber\\
&&\qquad\qquad
- \frac{Q_{l}(0)Q_{n}(0)\dot{Q}_{m}(0)}{\omega_{m}}
- \frac{Q_{l}(0)Q_{m}(0)\dot{Q}_{n}(0)}{\omega_{n}} \Bigg]
\sin((\omega_{l} - \omega_{m} - \omega_{n})t) 
\nonumber\\
&&\qquad\qquad
\Bigg[ \frac{\dot{Q}_{l}(0)\dot{Q}_{m}(0)\dot{Q}_{n}(0)}{\omega_{l}\omega_{m}\omega_{n}}
+ \frac{Q_{m}(0)Q_{n}(0)\dot{Q}_{l}(0)}{\omega_{l}} 
\nonumber\\
&&\qquad\qquad
- \frac{Q_{l}(0)Q_{n}(0)\dot{Q}_{m}(0)}{\omega_{m}}
+ \frac{Q_{l}(0)Q_{m}(0)\dot{Q}_{n}(0)}{\omega_{n}} \Bigg]
\sin((\omega_{l} - \omega_{m} + \omega_{n})t) 
\nonumber\\
&&\qquad\qquad
\Bigg[- \frac{\dot{Q}_{l}(0)\dot{Q}_{m}(0)\dot{Q}_{n}(0)}{\omega_{l}\omega_{m}\omega_{n}} 
+ \frac{Q_{m}(0)Q_{n}(0)\dot{Q}_{l}(0)}{\omega_{l}}
\nonumber\\
&&\qquad\qquad
 + \frac{Q_{l}(0)Q_{n}(0)\dot{Q}_{m}(0)}{\omega_{m}}
+ \frac{Q_{l}(0)Q_{m}(0)\dot{Q}_{n}(0)}{\omega_{n}} \Bigg]
\sin((\omega_{l} + \omega_{m} + \omega_{n})t) \Bigg\} \,,
\nonumber\\
&&
\end{eqnarray}
and using the results
\begin{equation}
\int_{0}^{t} \mbox{d}s \frac{ \cos(\alpha s) \sin(\omega_{k}(t-s))}{\omega_{k}} = 
\frac{ \cos(\omega_{k}t) - \cos(\alpha t)}{(\alpha + \omega_{k})(\alpha - \omega_{k})},
\end{equation}
and
\begin{equation}
\int_{0}^{t} \mbox{d}s \frac{ \sin(\alpha s) \sin(\omega_{k}(t-s))}{\omega_{k}} =
\frac{ \sin(\omega_{k}t) \alpha - \sin(\alpha t) \omega_{k}}{\omega_{k}(\alpha+\omega_{k})(\alpha-\omega_{k})}, 
\end{equation}
we see that the solution for $Q_{k,1}(t)$ is given by the following expression:
\begin{eqnarray}
 Q_{k,1}(t) &   & =  
\nonumber\\
&&
-\frac{\omega_{k}}{8(N+1)} \sum_{\langle l,m,n=1 \rangle}^{N} \omega_{l} \omega_{m} \omega_{n} C_{klmn}
\Bigg\{
   \Bigg[ Q_{l}(0)Q_{m}(0)Q_{n}(0) +\frac{Q_{l}(0)\dot{Q}_{m}(0)\dot{Q}_{n}(0)}{\omega_{m}\omega_{n}}
\nonumber\\
&& \qquad\qquad
 + \frac{Q_{m}(0)\dot{Q}_{l}(0)\dot{Q}_{n}(0)}{\omega_{l}\omega_{n}}
 - \frac{Q_{n}(0)\dot{Q}_{m}(0)\dot{Q}_{l}(0)}{\omega_{m}\omega_{l}}   \Bigg]
\left[\frac{ \cos(\omega_{k} t) - \cos(( \omega_{l} +  \omega_{m} -  \omega_{n}) t)}{
(\omega_{l} + \omega_{m} - \omega_{n}+\omega_{k})(\omega_{l}+\omega_{m}-\omega_{n}-\omega_{k})} \right]
\nonumber\\
&& \qquad\qquad
+
\Bigg[ Q_{l}(0)Q_{m}(0)Q_{n}(0) -\frac{Q_{l}(0)\dot{Q}_{m}(0)\dot{Q}_{n}(0)}{\omega_{m}\omega_{n}}
\nonumber\\
&& \qquad\qquad
 - \frac{Q_{m}(0)\dot{Q}_{l}(0)\dot{Q}_{n}(0)}{\omega_{l}\omega_{n}}
 - \frac{Q_{n}(0)\dot{Q}_{m}(0)\dot{Q}_{l}(0)}{\omega_{m}\omega_{l}} \Bigg]
\left[\frac{ \cos(\omega_{k} t) - \cos(( \omega_{l} +  \omega_{m} +  \omega_{n}) t)}{
(\omega_{l} + \omega_{m} + \omega_{n}+\omega_{k})(\omega_{l}+\omega_{m}+\omega_{n}-\omega_{k})} \right]
\nonumber\\
&& \qquad\qquad
+
\Bigg[ Q_{l}(0)Q_{m}(0)Q_{n}(0) -\frac{Q_{l}(0)\dot{Q}_{m}(0)\dot{Q}_{n}(0)}{\omega_{m}\omega_{n}}
\nonumber\\
&&\qquad\qquad
 + \frac{Q_{m}(0)\dot{Q}_{l}(0)\dot{Q}_{n}(0)}{\omega_{l}\omega_{n}}
 + \frac{Q_{n}(0)\dot{Q}_{m}(0)\dot{Q}_{l}(0)}{\omega_{m}\omega_{l}} \Bigg]
\left[\frac{ \cos(\omega_{k} t) - \cos(( \omega_{l} -  \omega_{m} -  \omega_{n}) t)}{
(\omega_{l} - \omega_{m} - \omega_{n}+\omega_{k})(\omega_{l}-\omega_{m}-\omega_{n}-\omega_{k})} \right]
\nonumber\\
&& \qquad\qquad
+
\Bigg[ Q_{l}(0)Q_{m}(0)Q_{n}(0) +\frac{Q_{l}(0)\dot{Q}_{m}(0)\dot{Q}_{n}(0)}{\omega_{m}\omega_{n}}
\nonumber\\
&&\qquad\qquad
 - \frac{Q_{m}(0)\dot{Q}_{l}(0)\dot{Q}_{n}(0)}{\omega_{l}\omega_{n}}
 + \frac{Q_{n}(0)\dot{Q}_{m}(0)\dot{Q}_{l}(0)}{\omega_{m}\omega_{l}} \Bigg] 
\left[\frac{ \cos(\omega_{k} t) - \cos(( \omega_{l} -  \omega_{m} +  \omega_{n}) t)}{
(\omega_{l} - \omega_{m} + \omega_{n}+\omega_{k})(\omega_{l}-\omega_{m}+\omega_{n}-\omega_{k})} \right]
\nonumber\\
&&\qquad\qquad
+
\Bigg[ \frac{\dot{Q}_{l}(0)\dot{Q}_{m}(0)\dot{Q}_{n}(0)}{\omega_{l}\omega_{m}\omega_{n}} 
+ \frac{Q_{m}(0)Q_{n}(0)\dot{Q}_{l}(0)}{\omega_{l}}
\nonumber\\
&&
 + \frac{Q_{l}(0)Q_{n}(0)\dot{Q}_{m}(0)}{\omega_{m}}
- \frac{Q_{l}(0)Q_{m}(0)\dot{Q}_{n}(0)}{\omega_{n}} \Bigg]
\left[\frac{ \sin(\omega_{k} t)(\omega_{l}+\omega_{m}-\omega_{n}) 
 -\sin((\omega_{l}+\omega_{m}-\omega_{n}) t)\omega_{k}}{
\omega_{k}(\omega_{l}+\omega_{m}-\omega_{n}+\omega_{k})(\omega_{l}+\omega_{m}-\omega_{n}-\omega_{k})} \right] 
\nonumber\\
&&\qquad\qquad
+
\Bigg[- \frac{\dot{Q}_{l}(0)\dot{Q}_{m}(0)\dot{Q}_{n}(0)}{\omega_{l}\omega_{m}\omega_{n}} 
+ \frac{Q_{m}(0)Q_{n}(0)\dot{Q}_{l}(0)}{\omega_{l}}
\nonumber\\
&&
 - \frac{Q_{l}(0)Q_{n}(0)\dot{Q}_{m}(0)}{\omega_{m}}
- \frac{Q_{l}(0)Q_{m}(0)\dot{Q}_{n}(0)}{\omega_{n}} \Bigg]
\left[\frac{ \sin(\omega_{k} t)(\omega_{l}-\omega_{m}-\omega_{n})
 -\sin((\omega_{l}-\omega_{m}-\omega_{n}) t)\omega_{k}}{
\omega_{k}(\omega_{l}-\omega_{m}-\omega_{n}+\omega_{k})(\omega_{l}-\omega_{m}-\omega_{n}-\omega_{k})} \right] 
\nonumber\\
&&\qquad\qquad
+
\Bigg[ \frac{\dot{Q}_{l}(0)\dot{Q}_{m}(0)\dot{Q}_{n}(0)}{\omega_{l}\omega_{m}\omega_{n}} 
+ \frac{Q_{m}(0)Q_{n}(0)\dot{Q}_{l}(0)}{\omega_{l}}
\nonumber\\
&&
 - \frac{Q_{l}(0)Q_{n}(0)\dot{Q}_{m}(0)}{\omega_{m}}
+ \frac{Q_{l}(0)Q_{m}(0)\dot{Q}_{n}(0)}{\omega_{n}} \Bigg]
\left[\frac{ \sin(\omega_{k} t)(\omega_{l}-\omega_{m}+\omega_{n})
 -\sin((\omega_{l}-\omega_{m}+\omega_{n}) t)\omega_{k}}{
\omega_{k}(\omega_{l}-\omega_{m}+\omega_{n}+\omega_{k})(\omega_{l}-\omega_{m}+\omega_{n}-\omega_{k})} \right]
\nonumber\\
&&\qquad\qquad
+
\Bigg[- \frac{\dot{Q}_{l}(0)\dot{Q}_{m}(0)\dot{Q}_{n}(0)}{\omega_{l}\omega_{m}\omega_{n}} 
+ \frac{Q_{m}(0)Q_{n}(0)\dot{Q}_{l}(0)}{\omega_{l}}
\nonumber\\
&&
\hspace*{-1cm}
 + \frac{Q_{l}(0)Q_{n}(0)\dot{Q}_{m}(0)}{\omega_{m}}
+ \frac{Q_{l}(0)Q_{m}(0)\dot{Q}_{n}(0)}{\omega_{n}} \Bigg]
\left[\frac{ \sin(\omega_{k} t)(\omega_{l}+\omega_{m}+\omega_{n})
 -\sin((\omega_{l}+\omega_{m}+\omega_{n}) t)\omega_{k}}{
\omega_{k}(\omega_{l}+\omega_{m}+\omega_{n}+\omega_{k})(\omega_{l}+\omega_{m}+\omega_{n}-\omega_{k})} \right] 
\Bigg\} \,,
\label{zwei} \\
&&
\end{eqnarray}
The above expression is the principal result of this work.
In the following sections, we apply 
this formalism for specific $N$ and specific initial conditions.

\section{Example for $N=2$}
The major advantage of the pertubative solutions is their generality; the full set of pertubative solutions to the
FPU-$\beta$ chain presented above with $N$ degrees of freedom has $ 2 N$ independent arbritrary constants specified by the initial conditions, as
is necessary to represent a fully general solution.
Solitons, discrete breathers, and other exotic nonlinear excitations
 that have been observed in
laboratory FPU-$\beta$ -like systems may then presumbably be constructed from the above general pertubative solution with appropriately chosen
initial conditions, given that it does in fact represent the fully general solution in the small $\epsilon$ limit.
In this section we present some general solutions with initial conditions chosen so as to describe accurately these nonlinear excitations.
To illustrate the method, we construct explicit solutions for the simplest non-trivial system,
the $N=2$ system.
The equations of motion are:
\begin{equation}
\ddot{Q}_{1} + Q_{1}  =  -\frac{1}{2} \epsilon Q_{1}^{3} - \frac{3}{2} \epsilon Q_{1} Q_{2}^{2}, 
\label{imp}
\end{equation}
and
\begin{equation}
\ddot{Q}_{2} + 3 Q_{2}  =  -\frac{9}{2} \epsilon Q_{2}^{3} - \frac{3}{2} \epsilon Q_{1}^{2} Q_{2}. 
\label{dimp}
\end{equation}

One easily finds that
\begin{eqnarray}
\beta_{1} &  = & 1 +   \frac{3}{16} \epsilon \left( Q_{1}^{2}(0) + P_{1}^{2}(0) \right) + 
\frac{3}{8} \epsilon \left( Q_{2}^{2}(0) + \frac{1}{3} P_{2}^{2}(0) \right) \\
\beta_{2} & = & 1 +  \frac{9}{16} \epsilon \left( Q_{2}^{2}(0) + \frac{1}{3} P_{2}^{2}(0) \right) + 
\frac{1}{8} \epsilon \left( Q_{1}^{2}(0) + P_{1}^{2}(0) \right)
\end{eqnarray}

As an example of random initial conditions, we consider the case $(Q_{1}(0) = 1/10, P_{1}(0) = 1/10,
Q_{2}(0)=1$ and $P_{2}(0)=0)$. 
We let $\epsilon = 1/10$.
These parameter choices result in values for $\rho_{1}$ and $\rho_{2}$ of
$303/808$ and $113/200$, respectively (with dimensions of $[\epsilon^{-1}]$).
With these values for the initial conditions and $\epsilon$, the explicit 
solutions are:
\begin{eqnarray}
Q_{1}(t) & = & 
1/10\,\cos \left( {\frac {8303}{4000}}\,\sin \left( 1/6\,\pi  \right) 
t \right) +{\frac {400}{8303}}\,\sin \left( {\frac {8303}{4000}}\,\sin
 \left( 1/6\,\pi  \right) t \right)  \left( \sin \left( 1/6\,\pi 
 \right)  \right) ^{-1} \nonumber \\
 & &  -{\frac {1}{2400}}\,{\frac { \left( \cos
 \left( t \right) -\cos \left(  \left( 2\,\sqrt {3}-1 \right) t
 \right)  \right) \sqrt {3}}{-2+2\,\sqrt {3}}}+{\frac {1}{320000}}\,
\cos \left( t \right) \nonumber  \\
 & &  -{\frac {1}{320000}}\,\cos \left( 3\,t \right) \nonumber \\
& &  -
{\frac {1}{1600}}\,{\frac { \left( \cos \left( t \right) -\cos \left( 
 \left( 2\,\sqrt {3}+1 \right) t \right)  \right) \sqrt {3}}{2\,\sqrt 
{3}+2}}+{\frac {1}{4800}}\,{\frac {\sqrt {3} \left( \cos \left( t
 \right) -\cos \left(  \left( 1-2\,\sqrt {3} \right) t \right) 
 \right) }{2-2\,\sqrt {3}}} \nonumber \\
& &  +{\frac {1}{2400}}\,{\frac { \left( \sin
 \left( t \right)  \left( 2\,\sqrt {3}-1 \right) -\sin \left(  \left( 
2\,\sqrt {3}-1 \right) t \right)  \right) \sqrt {3}}{-2+2\,\sqrt {3}}}
\nonumber \\
& & +{\frac {1}{4800}}\,{\frac {\sqrt {3} \left( \sin \left( t \right) 
 \left( 1-2\,\sqrt {3} \right) -\sin \left(  \left( 1-2\,\sqrt {3}
 \right) t \right)  \right) }{2-2\,\sqrt {3}}}-{\frac {3}{320000}}\,
\sin \left( t \right) +{\frac {1}{320000}}\,\sin \left( 3\,t \right)
\nonumber \\
& &  -
{\frac {1}{1600}}\,{\frac { \left( \sin \left( t \right)  \left( 2\,
\sqrt {3}+1 \right) -\sin \left(  \left( 2\,\sqrt {3}+1 \right) t
 \right)  \right) \sqrt {3}}{2\,\sqrt {3}+2}}
\label{einer}
\end{eqnarray}
 
and
\begin{eqnarray}
Q_{2}(t) & = & 
\cos \left( {\frac {2113}{1000}}\,\sin \left( 1/3\,\pi  \right) t
 \right) -{\frac {3}{640}}\,\cos \left( \sqrt {3}t \right) +{\frac {3}
{640}}\,\cos \left( 3\,\sqrt {3}t \right) \nonumber \\
& &  -{\frac {1}{12000}}\,{\frac 
{\sqrt {3} \left( \sin \left( \sqrt {3}t \right)  \left( 2-\sqrt {3}
 \right) -\sin \left(  \left( 2-\sqrt {3} \right) t \right) \sqrt {3}
 \right) }{2-2\,\sqrt {3}}} \nonumber \\
& & -{\frac {1}{24000}}\,{\frac {\sqrt {3}
 \left( \sin \left( \sqrt {3}t \right)  \left( \sqrt {3}-2 \right) -
\sin \left(  \left( \sqrt {3}-2 \right) t \right) \sqrt {3} \right) }{
-2+2\,\sqrt {3}}} \nonumber \\
& & -{\frac {1}{8000}}\,{\frac {\sqrt {3} \left( \sin
 \left( \sqrt {3}t \right)  \left( \sqrt {3}+2 \right) -\sin \left( 
 \left( \sqrt {3}+2 \right) t \right) \sqrt {3} \right) }{2\,\sqrt {3}
+2}}
\label{ander}
\end{eqnarray}

\begin{figure}
\psfig{figure=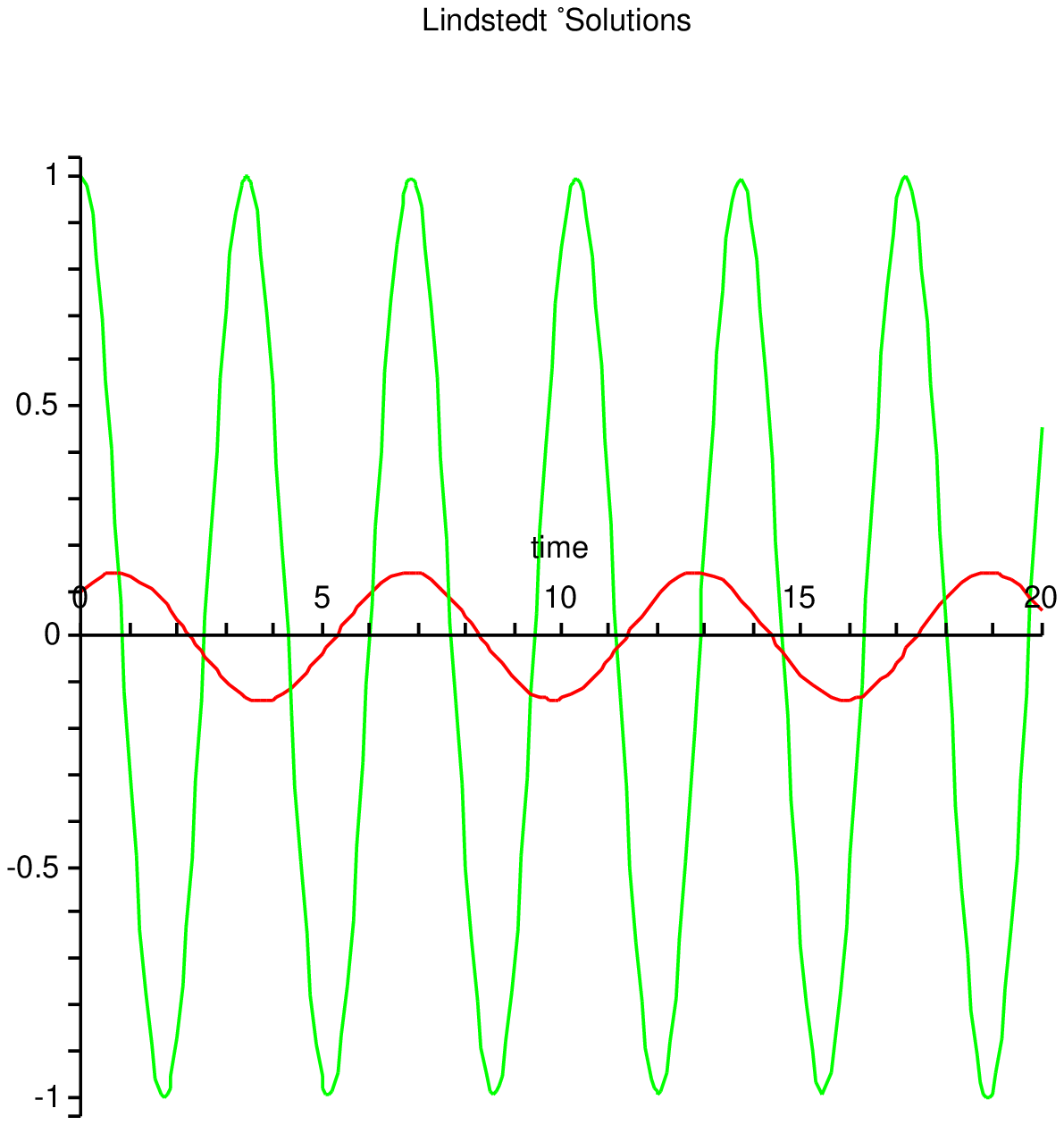,width=12cm,height=10cm}
\caption{$Q_{1}(t), Q_{2}(t)$ for the initial conditions given in the text. $Q_{1}(t)$ is 
represented by the lower-amplitude curve.}
\label{lind}
\end{figure}

\begin{figure}
\psfig{figure=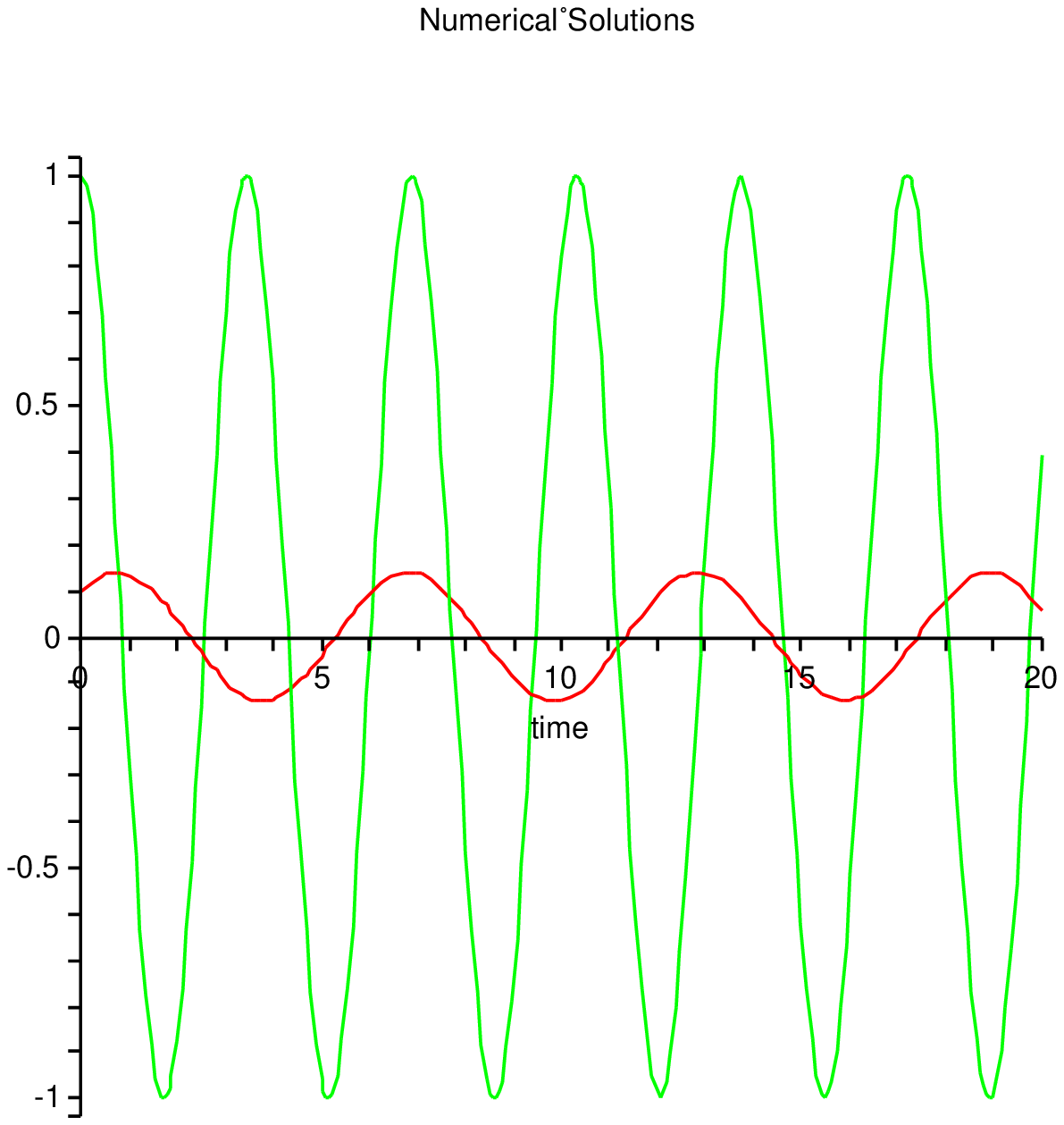,width=12cm,height=10cm}
\caption{Numerical solutions for $Q_{1}(t)$ and $Q_{2}(t)$ for the initial conditions given in the text. 
$Q_{1}(t)$ is represented by the lower-amplitude curve.}
\label{zahleric}
\end{figure}

In Fig.~(\ref{lind}), we plot $Q_{1}(t)$ and $Q_{2}(t)$ and in Fig.~(\ref{zahleric}), we plot 
the numeric solutions for the same system.
\nopagebreak
\nopagebreak

\begin{figure}
\psfig{figure=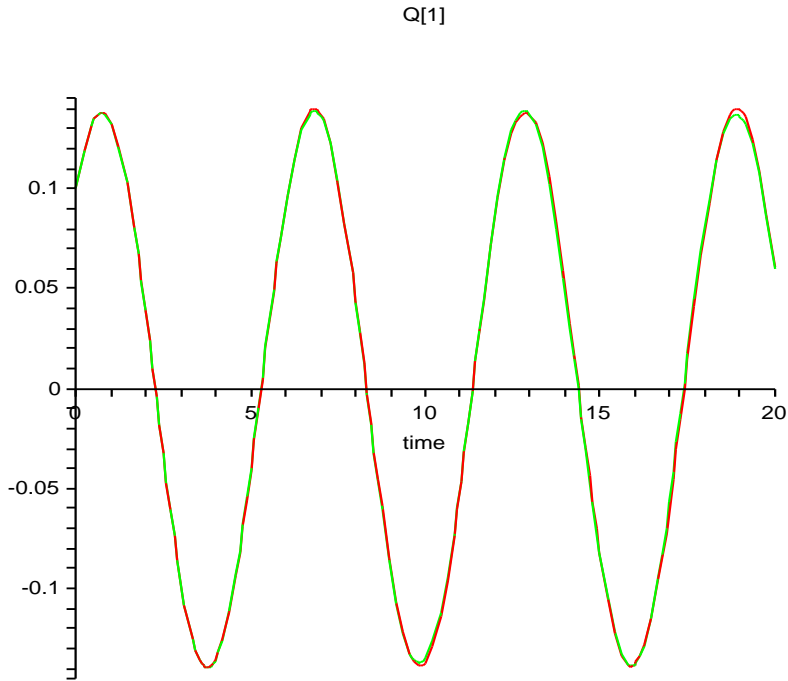,width=12cm,height=10cm}
\caption{Numerical solutions for $Q_{1}(t)$ and the analytical approximation for $Q_{1}(t)$ for the initial conditions given in the text.}
\label{coord1}
\end{figure}

\begin{figure}
\psfig{figure=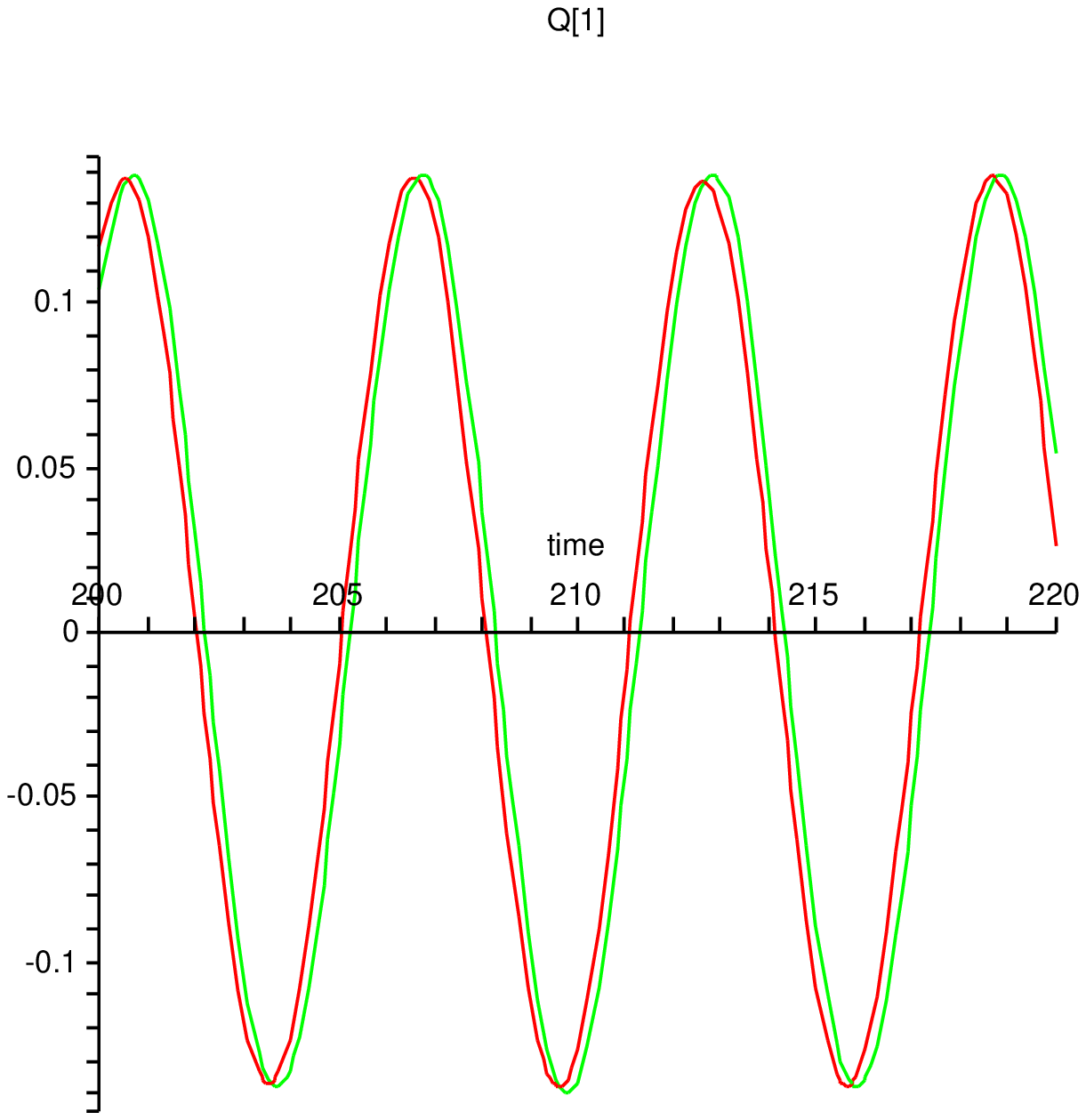,width=12cm,height=10cm}
\caption{Numerical solutions for $Q_{1}(t)$ and the analytical approximation for $Q_{1}(t)$ for the initial conditions given in the \
text.}
\label{1q200}
\end{figure}

\begin{figure}
\psfig{figure=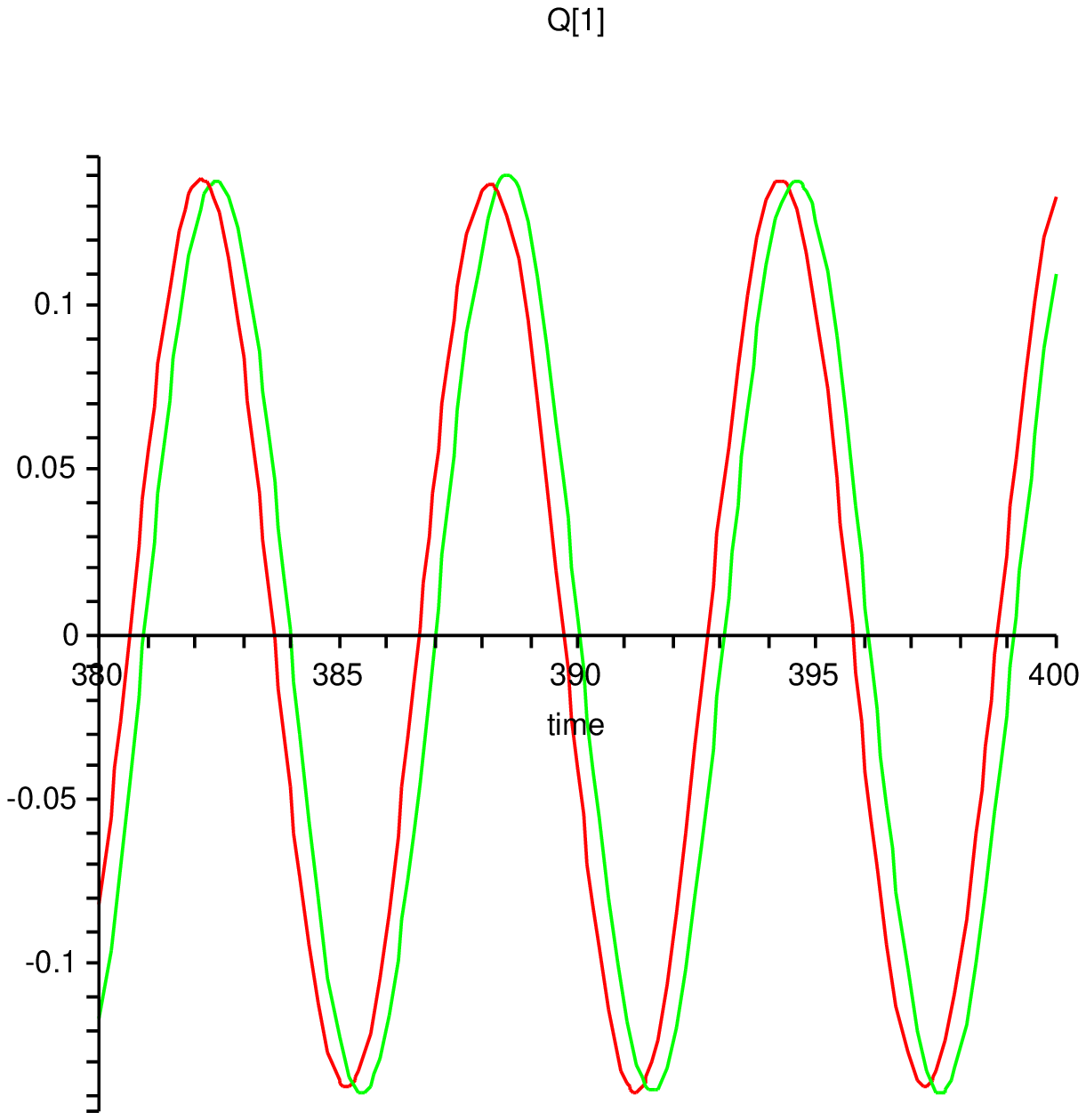,width=12cm,height=10cm}
\caption{Numerical solutions for $Q_{1}(t)$ and the analytical approximation for $Q_{1}(t)$ for the initial conditions given in the \
text.}
\label{1q400}
\end{figure}

\begin{figure}
\psfig{figure=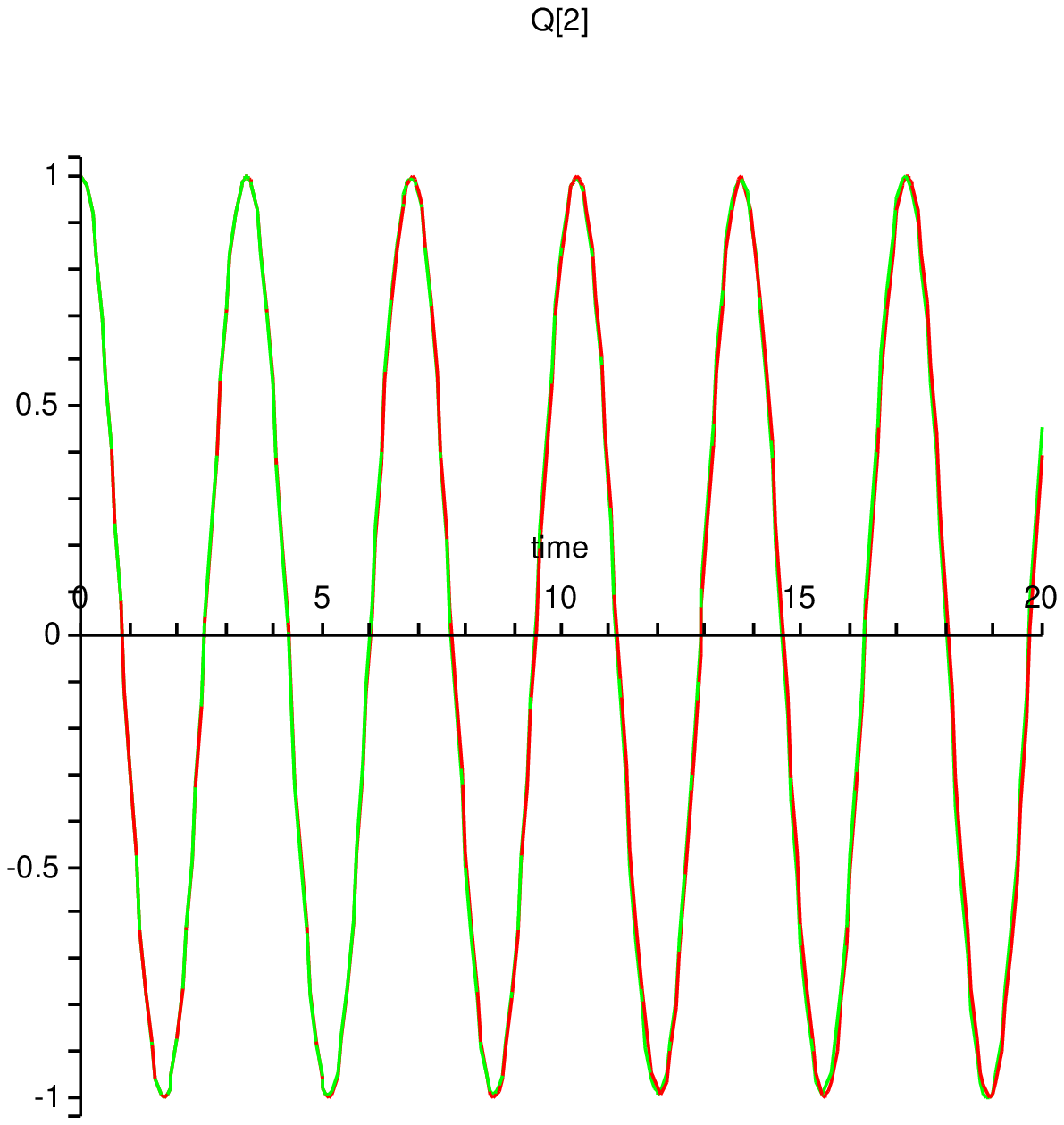,width=12cm,height=10cm}
\caption{Numerical solutions for $Q_{2}(t)$ and the analytical approximation for $Q_{2}(t)$ for the initial conditions given in the text.}
\label{coord2}
\end{figure}

\begin{figure}
\psfig{figure=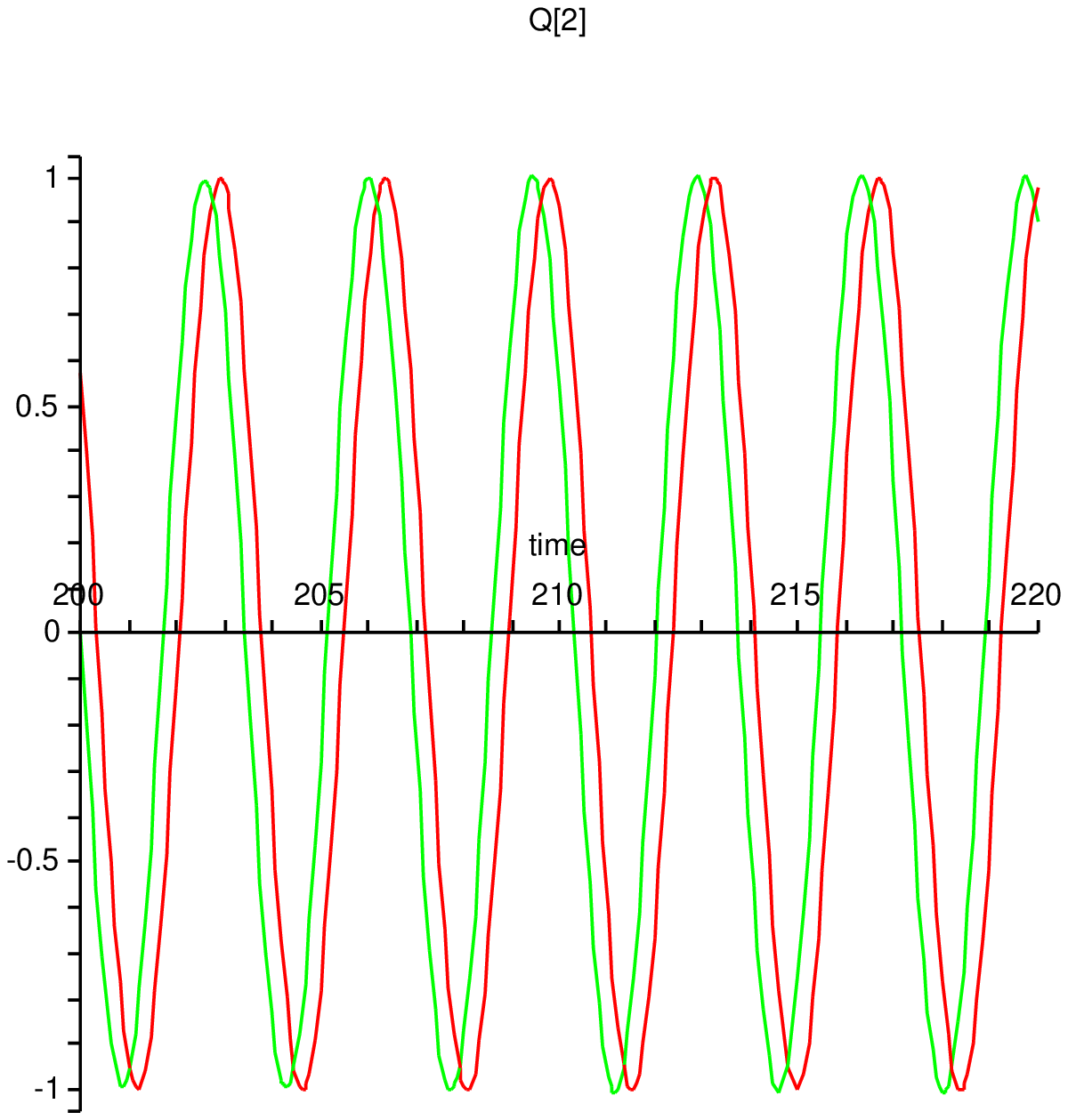,width=12cm,height=10cm}
\caption{Numerical solutions for $Q_{2}(t)$ and the analytical approximation for $Q_{2}(t)$ for the initial conditions given in the text.}
\label{2q200}
\end{figure}

\begin{figure}
\psfig{figure=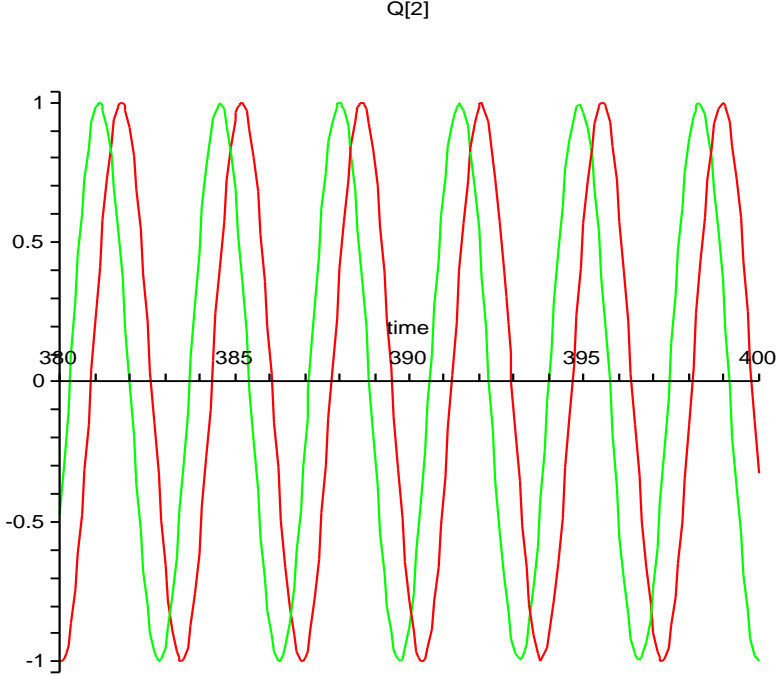,width=12cm,height=10cm}
\caption{Numerical solutions for $Q_{2}(t)$ and the analytical approximation for $Q_{2}(t)$ for the initial conditions given in the text.}
\label{2q400}
\end{figure}

In Figs.~(\ref{coord1},\ref{1q200}, \ref{1q400}), we plot both the numerical solution and the analytical approximation for $Q_{1}$ for
three different time intervals of twenty units, and
In Figs.~(\ref{coord2}, \ref{2q200}, \ref{2q400}), we plot both the numerical solution and the analytical approximation for $Q_{2}$ for
the same time intervals.

\begin{figure}
\psfig{figure=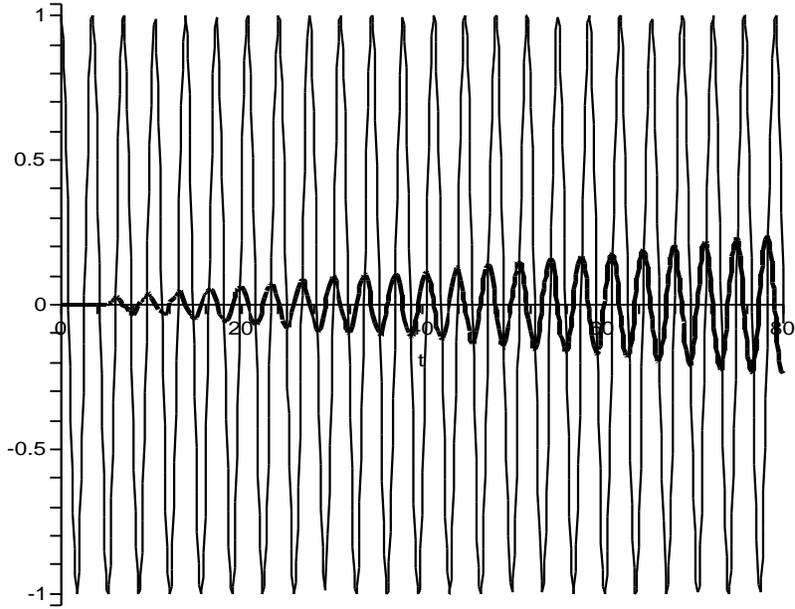,width=12cm,height=10cm}
\caption{Plots of $Q_{2,\mbox{num}}(t)$ (thin curve) and
Eqn.~(\ref{dork}) (thick curve) .} 
\label{time80}
\end{figure}

 In Fig.~(\ref{time80}) we plot
\begin{equation}
Q_{2,0}(\tau_{2}=\beta_{2} t) + \epsilon Q_{2,1}(t) - Q_{2,\mbox{num}}(t)
\label{dork}
\end{equation}
and
$Q_{2,\mbox{num}}(t)$,
, where
$Q_{2,\mbox{num}}(t)$ is the numerical solution for $Q_{2}(t)$ of the system of 
eqns.~(\ref{imp},\ref{dimp}).

\begin{figure}
\psfig{figure=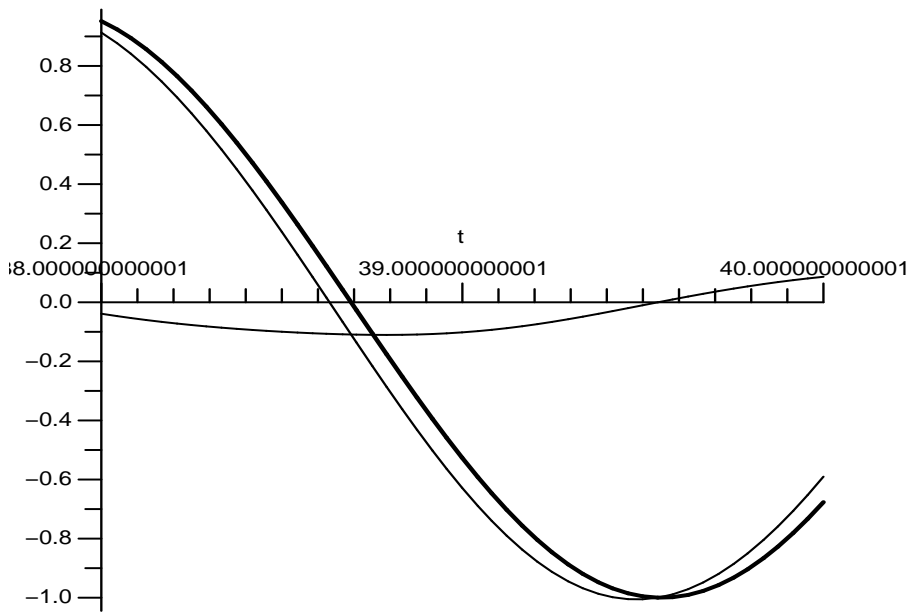,width=12cm,height=10cm}
\caption{Plots of
 $(Q_{2,\mbox{num}}(t)$ (thick curve),
our perturbative approximation given by Eqn.~(\ref{ander}) (thin curve), and 
the their difference, Eqn.~(\ref{dork}). } 
\label{timeyo}
\end{figure}

In Fig.~(\ref{timeyo}), we plot the quantities $Q_{2,\mbox{num}}(t)$ (black curve),
our perturbative approximation given by Eqn.~(\ref{ander}) (grey curve), and 
the their difference, Eqn.~(\ref{dork}).
From Fig.~(\ref{timeyo}), we note that the principal source of the 
growing disagreement between our approximate solution and the numerical
solution arises from the difference in phases between the two functions.
The amplitudes appear to be in strong agreement.
Furthermore, from Fig.~(\ref{time80}), it appears 
that the phase difference is some linear function of time.
We thus speculate that the full numerical solution has not only
an 
amplitude-dependent frequency, but also a time-dependent frequency.
If we allow for time-dependent frequency shifts $\rho_{k}$,
we may achieve greater agreement for much longer times.
We leave this open question for a further investigation.

\section{Conclusion}
The FPU-$\beta$ lattice has been a subject of study in relation to fundamental issues in nonlinear 
dynamical systems since its introduction.
Because a single Duffing oscillator admits of an exact solution, one may be tempted to entertain the notion that a lattice
of Duffing oscillators with nearest neighbor coupling may also allow fully general analytic solutions.
For special initial conditions, the full lattice dynamics is described by a single quartic oscillator and thus admits 
exact analytic solutions \cite{Poggi,Rink}, but these particular solutions correspond to special initial conditions.
While for a small number $N$ of degrees of freedom, some of these special solutions are stable over an
appreciable range of $\epsilon$, they are actually unstable for arbitrarily small $\epsilon$
in the limit of $N$ approaching infinity. 
For more general initial conditions, we have applied the Lindstedt method to lowest order in the nonlinearity
parameter $\epsilon$ to obtain analytical, quasiperiodic approximations. 
These quasiperiodic solutions are predicted to exist on the basis of the KAM (Kolmogorov, Arnold and Moser) theorem and is the
accepted explanation for the nonchaotic behavior and recurrences that puzzled Fermi, Pasta and Ulam.
As noted above, the formal perturbative scheme presented above has unknown convergence properties. 
For small $\epsilon$ and early times, the solutions of the formalism and numerically obtained solutions are
 numerically indistinguishable.
For the example in the text, the pertubative and numerical solutions begin to noticeably differ in phase for dimensionless time 
on the order of forty.
We conclude that
the formal perturbation expansions developed here, truncated at first order in $\epsilon$, give useful
analytic approximations to quasi-periodic solutions of the FPU-$\beta$ lattice for
moderately long times.
Finally, we remark that the general application of the above perturbative scheme is independent of the spatial
dimension of the system.

\end{document}